\documentclass[fleqn,twoside]{article}
\usepackage{espcrc2}

\usepackage{graphicx}
\usepackage[figuresright]{rotating}
\RequirePackage{xspace}

\def\CP     {\ensuremath{C\!P}\xspace}
\def\Bbar   {\kern 0.18em\overline{\kern -0.18em B}{}\xspace}
\def\Dbar   {\kern 0.2em\overline{\kern -0.2em D}{}\xspace}
\def\Db     {\ensuremath{\Dbar}\xspace}
\def\jpsi   {\ensuremath{{J\mskip -3mu/\mskip -2mu\psi\mskip 2mu}}\xspace}
\def\psitwos{\ensuremath{\psi{(2S)}}\xspace}
\def\KS     {\ensuremath{K^0_{\scriptscriptstyle S}}\xspace} 
\def\KL     {\ensuremath{K^0_{\scriptscriptstyle L}}\xspace} 
\mathchardef\Upsilon="7107
\def\Y#1S{\ensuremath{\Upsilon{(#1S)}}\xspace}
\def\FourS  {\Y4S}

\newcommand {\gevcc}{\ensuremath{{\mathrm{\,Ge\kern -0.1em V\!/}c^2}}\xspace}
\newcommand {\tevcc}{\ensuremath{{\mathrm{\,Te\kern -0.1em V\!/}c^2}}\xspace}
\def\cm     {\ensuremath{{\rm \,cm}}\xspace}
\def\sec    {\ensuremath{\rm {\,s}}\xspace}

\newcommand{\stat}{\ensuremath{\mathrm{(stat)}}\xspace}
\newcommand{\syst}{\ensuremath{\mathrm{(syst)}}\xspace}

\title{Results on $\CP$ violation and CKM UT angles from Belle and BaBar}

\author{Gagan B. Mohanty\address{Tata Institute of Fundamental Research, 
        Homi Bhabha Road, Mumbai 400005, India}\,%
        \thanks{Tel.: +91\,22\,22782147; Fax: +91\,22\,22804610;
        E-mail: gmohanty@tifr.res.in}}
       
\begin{document}

\begin{abstract}
We report recent results on $\CP$ violation measurements from the two
$B$-factory experiments, Belle and BaBar.
\vspace{1pc}
\end{abstract}

\maketitle

\section{INTRODUCTION}

In the standard model (SM), $\CP$ violation occurs due to a single,
irreducible phase appearing in the $3\times 3$ quark-flavor mixing
matrix, called the Cabibbo-Kobayashi-Maskawa (CKM) matrix \cite{ckm},
which relates quark mass eigenstates to weak eigenstates. Unitarity
of the CKM matrix yields a set of relations among its elements that
can be depicted as triangles in the complex plane. In particular, the
unitarity condition $V_{ud}V^*_{ub}+V_{cd}V^*_{cb}+V_{td}V^*_{tb}=0$
gives rise to the so-called unitarity triangle (UT), whose sides and
angles are related to the magnitudes and phases of the CKM matrix
elements $V_{id}$ and $V_{ib}$, where $i=u,c,t$. The main goal of the
two $B$-factory experiments -- Belle \cite{belle} at KEK, Japan and
BaBar \cite{babar} at SLAC, USA -- is to overconstrain the UT through
precise measurements of its sides and angles. By doing so, they are
designed to verify whether the CKM mechanism is the correct description
of $\CP$ violation in the SM, and to set constraints on possible new
physics effects that could lead to inconsistencies among these measurements.

In these proceedings, we summarize recent results on $\CP$ violation,
involving three UT angles, from Belle and BaBar. After a decade of
successful operation, during which many records are made and broken
subsequently, these two $B$-factory experiments have together collected
over $10^9$ $B\Bbar$ pairs at the $\FourS$ peak. The KEKB accelerator
of the $B$ factory in Japan holds the current world record with a peak
luminosity $2.1\times10^{34}\cm^{-2}\sec^{-1}$. Results reported here
comprise the full $\FourS$ data from BaBar ($\sim 465\times 10^6$ $B\Bbar$)
and a large fraction of the $\FourS$ data available with Belle ($\sim
535\times 10^6$ $B\Bbar$). 

\section{ANGLES OF THE UNITARITY TRIANGLE}

The UT angles are mostly determined through the measurement of the
time-dependent \CP\ asymmetry,
\begin{equation}
A_{\CP}(t) = \frac{N[\Bbar^0(t)\to f_{\CP}]-N[B^0(t)\to f_{\CP}]}
                  {N[\Bbar^0(t)\to f_{\CP}]+N[B^0(t)\to f_{\CP}]},
\label{eq1}
\end{equation}
where $N[\Bbar^0(t)/B^0(t)\to f_{\CP}]$ is the number of $\Bbar^0/B^0$s
that decay into a common \CP\ eigenstate $f_{\CP}$ after time $t$. The
asymmetry, in general, can be expressed as
\begin{equation}
A_{\CP}(t)=S_f\sin(\Delta mt)+A_f\cos(\Delta mt),
\label{eq2}
\end{equation}
where $\Delta m$ is the mass difference between the two $B^0$ mass
eigenstates. (Note that BaBar uses a notation $C_f= -A_f$.) The sine
coefficient $S_f$ here is related to the UT angles, while the cosine
coefficient $A_f$ is a measure of direct \CP\ violation. For the latter
to have a nonzero value, one needs at least two competing amplitudes with
different weak and strong phases to contribute to the decay final state.
As an example, for the decay $B^0\to\jpsi\KS$, where mostly one diagram
contributes, the cosine term is expected to vanish and the sine term is
proportional to the UT angle $\phi_1$\footnote{An alternative notation
of $\beta$, $\alpha$, and $\gamma$ corresponding to $\phi_1$, $\phi_2$,
and $\phi_3$, respectively, is adopted by BaBar.}. The time-dependent
$\CP$ asymmetry is, therefore, given as
\begin{equation}
A_{\CP}(t) = -\xi_f\sin(2\phi_1)\sin(\Delta mt),
\label{eq3}
\end{equation}
where $\xi_f$ is the $\CP$ eigenvalue of the final state $f_{\CP}$.
In the case of $B$ factories, the measurement of $A_{\CP}(t)$ utilizes
decays of the \FourS\ into two neutral $B$ mesons, of which one can be
fully reconstructed into a \CP\ eigenstate, while decay products of the
other (called the tag $B$) identify its flavor at the decay time. The time
difference $t$ between the two $B$ decays is determined by reconstructing
their decay vertices. Finally the $\CP$ asymmetry amplitudes, proportional
to the UT angles, are obtained from a maximum likelihood fit to the proper
time distributions separately for events tagged as $\Bbar^0$ and $B^0$.

\subsection{The angle {\boldmath $\phi_1$}}

The most precise measurement of the angle $\phi_1$ is obtained from a
study of the decays $B^0\to$ charmonium $+\,K^{(*)0}$. These decays, known
as ``golden" modes, mainly proceed via the CKM-favored tree diagram
$b\to c\bar{c}s$ with an internal $W$-boson emission. The subleading
penguin (loop) contribution to the final state, having a different weak
phase compared to the tree diagram, is suppressed by almost two orders of
magnitude. This makes $A_f=0$ in Eq.~\ref{eq2} to a good approximation.
Besides the theoretical simplicity, these channels also offer experimental
advantages because of the relatively large branching fractions ($\sim
10^{-3}$) and the presence of narrow resonances in the final state, which
provides a powerful rejection against the combinatorial background. The
\CP\ eigenstates considered for this analysis include $\jpsi\KS$,
$\psitwos\KS$, $\chi_{c0}\KS$, $\eta_c\KS$, and $\jpsi\KL$.

\begin{figure}[!htb]
\center
\includegraphics[width=\columnwidth]{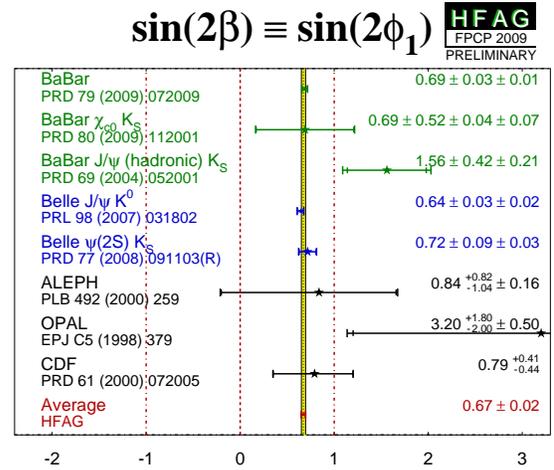}
\caption{Average of $\sin(2\phi_1)$ from all experiments, as compiled
         by the HFAG.}
\label{sin2phi1}
\end{figure}

BaBar has updated the $\sin(2\phi_1)$ measurement with its full $\FourS$
data sample \cite{babarphi1}. The result is $\sin(2\phi_1)=0.687\pm 0.028
\stat\pm 0.012\syst$. Combined with Belle's result based on $535\times 10^6$
$B\Bbar$ pairs \cite{bellephi1}, $\sin(2\phi_1)=0.642\pm 0.031\stat\pm 0.017
\syst$, the world-average value is $\sin(2\phi_1)= 0.672\pm 0.023$ \cite{HFAG}.
The result, having a precision of $3\%$, serves as a firm reference point
for the SM. The world-average value of $A_f$ is $0.004\pm 0.019$ \cite{HFAG},
which is consistent with zero as expected. This major accomplishment of the $B$
factories has been cited \cite{nobel} as leading to half of the 2008 physics
Nobel prize being awarded to Kobayashi and Maskawa. Figure \ref{sin2phi1}
compares the impact of measurements from Belle and BaBar with those from the
other experiments.

\subsection{The angle {\boldmath $\phi_2$}}

Decays of $B$ mesons to the final states $hh$ ($h=\rho$ or $\pi$),
dominated by the CKM-suppressed $b\to u$ transition, are sensitive
to the angle $\phi_2$. The presence of $b\to d$ penguin diagrams,
however, complicates the situation by introducing additional phases
such that the measured parameter is no more $\phi_2$ alone, rather
an effective value $\phi^{\rm eff}_2 =\phi_2+\delta\phi_2$. Through
an isospin analysis \cite{gronau1} one can isolate the tree
contribution, and hence the $\phi_2$ value. At present, the most
precise measurement of this angle is obtained in the analysis of
the decays $B\to\rho\rho$. Combining with additional constraints
coming from $B\to\rho\pi$ and $B\to\pi\pi$, we measure $\phi_2=
\left(89.0^{+4.4}_{-4.2}\right)^\circ$ \cite{ckmfitter}.

\subsection{The angle {\boldmath $\phi_3$}}

\begin{figure*}[!htb]
\center
\includegraphics[width=1.45\columnwidth]{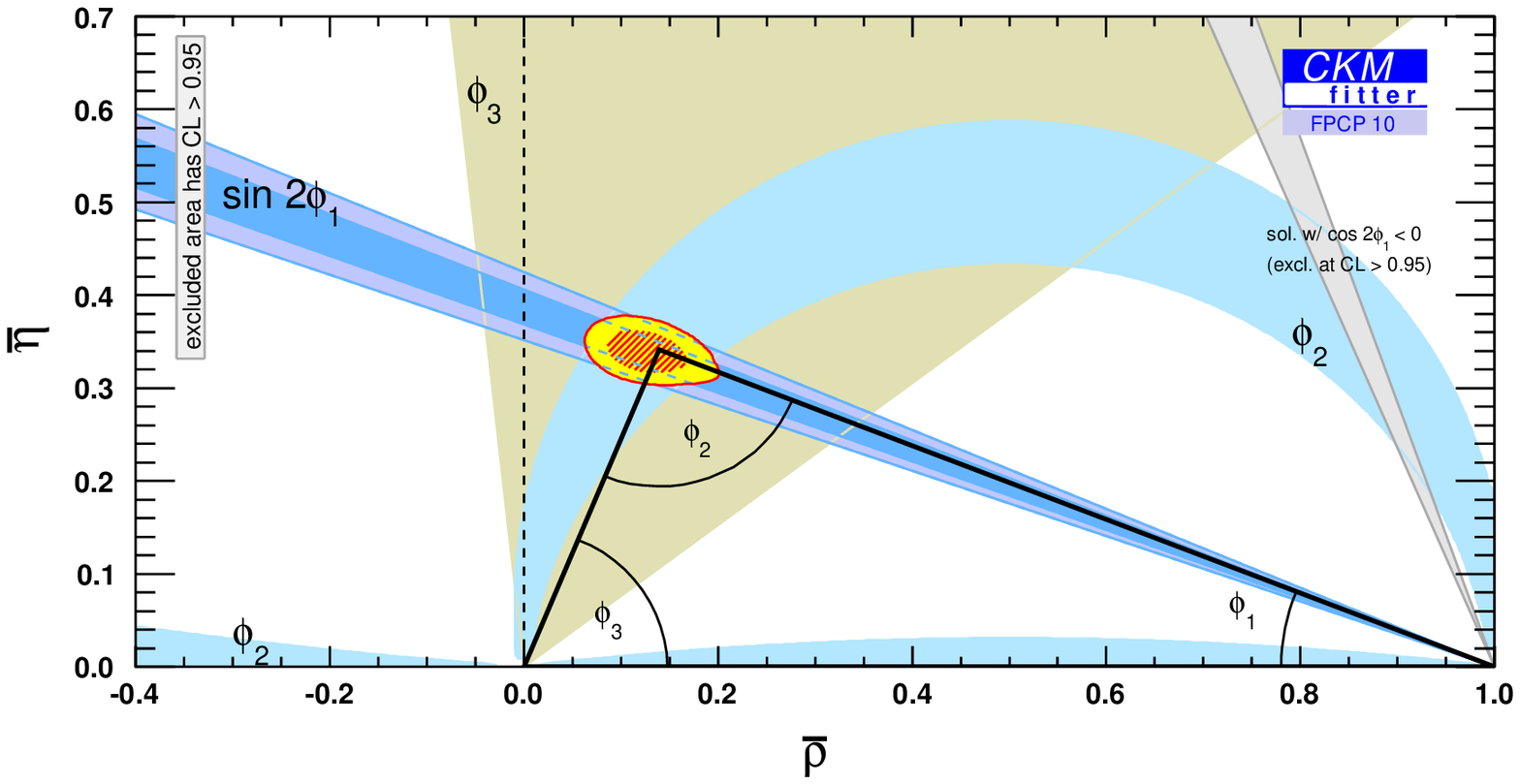}\\
\includegraphics[width=1.45\columnwidth]{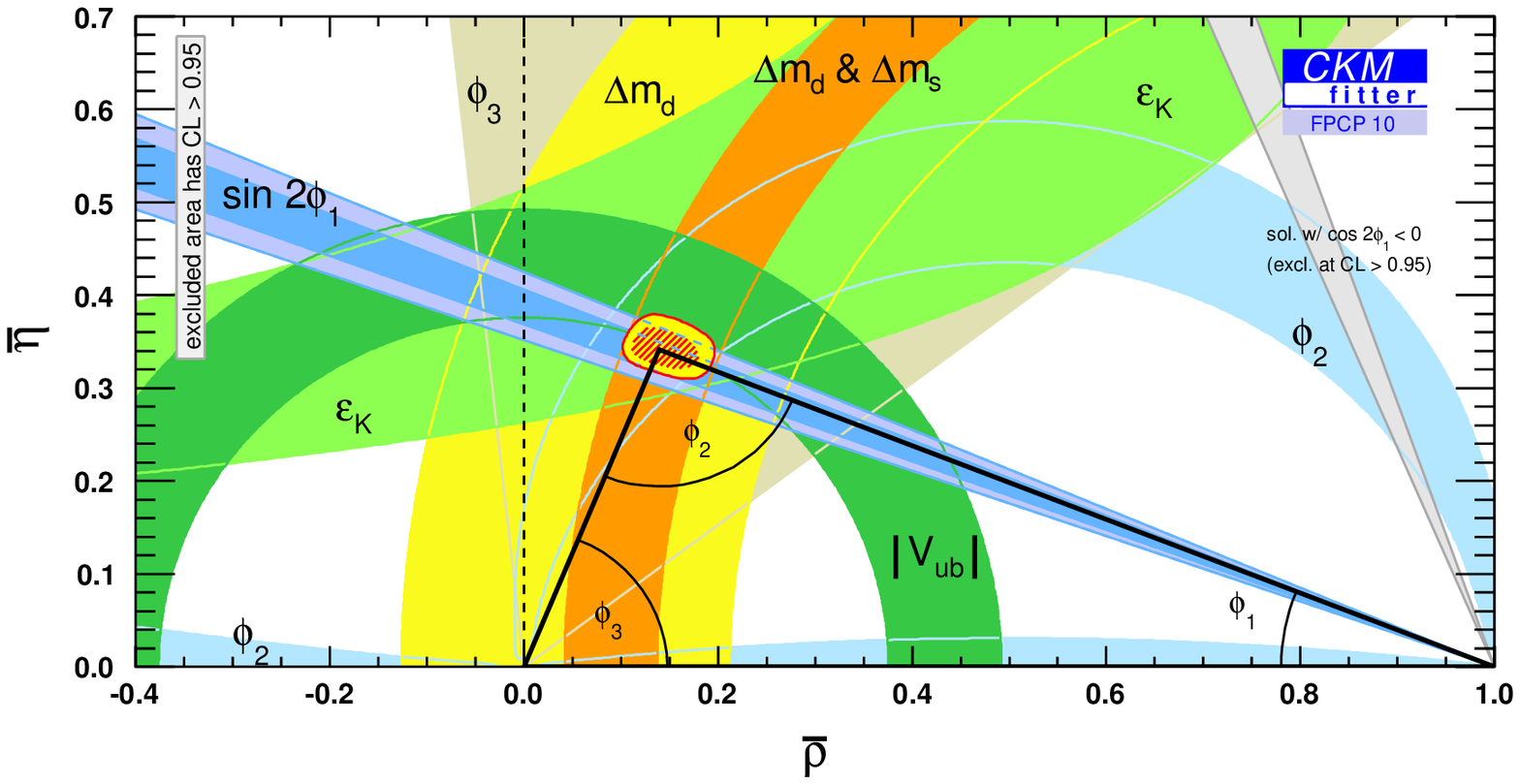}
\caption{Constraints on the UT \cite{ckmfitter} coming from the
         measurements of angles only (above) and using all relevant
         experimental inputs (below).}
\label{ckmfitresult}
\end{figure*}
The angle $\phi_3$ cannot be extracted using time-dependent $\CP$
violation study in a similar fashion as was done for other two
angles. It is rather measured by exploiting the interference between
$B^-\to D^{(*)0}K^{(*)-}$ (dominated by the $b\to c$ tree diagram
with an external $W$ emission) and $B^-\to\Db^{(*)0}K^{(*)-}$
(dominated by the color-suppressed $b\to u$ tree diagram with an
internal $W$ emission). Here, both $D^0$ and $\Dbar^0$ decay to a
common final state. This measurement can be performed in three
different ways: (a) by utilizing decays of $D$ mesons to $\CP$
eigenstates, such as $\pi^+\pi^-$, $K^+K^-$ ($\CP$ even) or
$\KS\pi^0$, $\phi\KS$ ($\CP$ odd) \cite{glw}, (b) by making use
of doubly Cabibbo suppressed decays of $D$ mesons, {\it e.g.},
$D^0\to K^+\pi^-$ \cite{ads}, or (c) by exploiting the interference
pattern in the Dalitz plot of the decays $D\to\KS\pi^+\pi^-$
\cite{ggsz}. The first two methods are theoretically clean but
suffer from low statistics. On the other hand, the Dalitz method
currently provides the strongest constraint on $\phi_3$. Combining
all recent measurements from Belle \cite{bellephi3} and BaBar
\cite{babarphi3}, the world-average value is found to be $\phi_3=
\left(70^{+14}_{-21}\right)^\circ$ \cite{ckmfitter}.

\subsection{Putting them together}

In Fig.~\ref{ckmfitresult} we summarize constraints on the UT
coming from the measurements of angles only, as well as after
including other experimental inputs. To a good approximation,
the CKM framework is found to be the right description of $\CP$
violation in the SM. Needless to say that the precision on the
third angle $\phi_3$ ought to be improved. Similarly, we expect
errors on the other two angles to shrink further, {\it e.g.},
once Belle analyzes its full $\FourS$ dataset.

\subsection{Probing new physics in {\boldmath $\CP$} violation}

\begin{figure}[!htb]
\center
\includegraphics[width=\columnwidth]{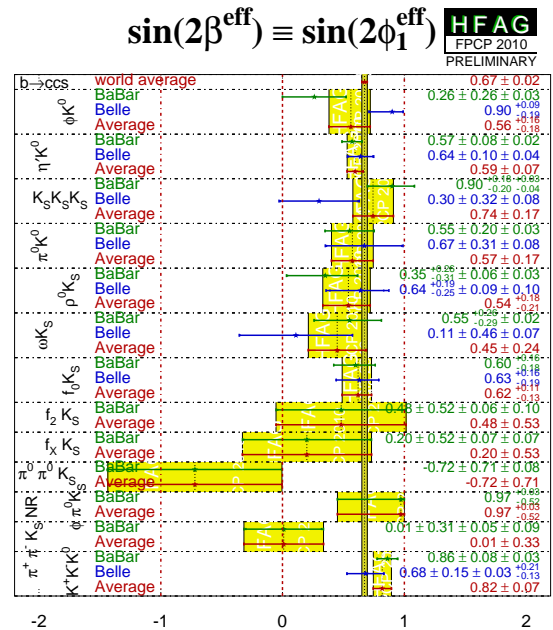}
\caption{Time-dependent $\CP$ asymmetry measured in the $b\to s$
 penguin decay channels \cite{HFAG}.}
\label{penguin}
\end{figure}

As $\sin(2\phi_1)$ is the most precisely measured observable
concerning $\CP$ violation in $B$ decays, one can use it as a
``Standard Candle'' to set constraints on new physics by looking for
possible deviations from this value in a number of ways. One such is
the comparison of the values of $\sin(2\phi^{\rm eff}_1)$ measured
in penguin dominated decays with the world-average value of
$\sin(2\phi_1)$, coming from decays involving charmonium final states.
The results are summarized in Fig.~\ref{penguin}, where the largest
discrepancy is found to be at the level of $2$ standard deviations. A
caveat one should be aware of while making such a comparison is that
the penguin modes may have additional topologies that could lead to a
difference between $\sin(2\phi_1)$ and $\sin(2\phi^{\rm eff}_1)$. If
these SM corrections, $\Delta_{\rm SM}$, are well known then any
residual difference $\Delta S=\sin(2\phi^{\rm eff}_1)-\sin(2\phi_1)-
\Delta_{\rm SM}$ would be from new physics. Nevertheless, looking at
Fig.~\ref{penguin} it is fair to say that we need more data before
drawing a firm conclusion whether the observed deviations are due
to some new physics effects or a play of statistics.

\section{DIRECT {\boldmath $\CP$} VIOLATION}

Both Belle and BaBar have intensively searched for direct $\CP$
violation in several $B$ decays. The most notable result comes
from the decay $B^0\to K^+\pi^-$, where direct $\CP$ violation
has been established beyond any doubt: the measured $\CP$
asymmetry is $\left(-9.8^{+1.2}_{-1.1}\right)\%$. This is in
contrast to the result from $B^-\to K^-\pi^0$ having a $\CP$
asymmetry $(+5.0\pm 2.5)\%$. Since both the decays are expected
to proceed via similar Feynman diagrams at the tree level, the
discrepancy between the two measurements \cite{nature} tells
us that it could be either due to a large contribution from the
color-suppressed tree diagram, or from possible new physics
contribution in the electroweak penguin, or from a mixture of both.
Before concluding anything, it has been suggested~\cite{sumrule}
to improve the precision on $\CP$ violation results of the decay
$B^0\to K^0\pi^0$ \cite{kspi0} using more data. In addition to
these results, there are also a number of interesting evidences for
direct $\CP$ violation at a level of $3$ standard deviations in the
decays $B^0\to\eta K^{*0}$, $B^-\to\eta K^-$, $B^-\to\rho^0 K^-$,
$B^0\to\rho^+\pi^-$, $B^-\to f_2(1270)K^-$ and $B^-\to\Db^{(*)0}K^-$.

\section{SUMMARY}

Thanks to the excellent performance of the two $B$ factories,
studies using a large sample of $e^+e^-$ collision data at the
$\FourS$ peak have now established the CKM framework as the
only source of $\CP$ violation in the SM. There are a number of
intriguing hints, such as time-dependent $\CP$ asymmetry in
penguin dominated decays and direct $\CP$ asymmetry difference
in $B\to K\pi$, at various levels of significance. These results
need to be clarified with much larger data samples. Towards this
end, we are eagerly looking forward to final updates from Belle
in several important channels, {\it e.g.} the angle $\phi_1$ in
the $B\to$ charmonium $+\,K^{(*)0}$ decays, while warming up to
the next generation of flavor experiments: LHCb and super flavor
factories.

\section{ACKNOWLEDGEMENTS}

I thank my Belle colleagues for their valuable helps in the
preparation of these proceedings. The work is supported in
parts by the Department of Atomic Energy and the Department
of Science and Technology of India.

\end{document}